\documentclass[twocolumn,showpacs,preprintnumbers,amsmath,amssymb]{revtex4}

\usepackage{graphicx}
\usepackage{dcolumn}
\usepackage{bm}

\begin{document}

\title{Quantum entanglement of decohered two-mode squeezed states
in absorbing and amplifying environment}

\author{Phoenix S. Y. Poon and C. K. Law}
\affiliation{Department of Physics and Institute of Theoretical
Physics, The Chinese University of Hong Kong, Shatin, Hong Kong
SAR, China}

\date{\today}
\begin{abstract}
We investigate the properties of quantum entanglement of two-mode
squeezed states interacting with linear baths with general gain and
loss parameters. By explicitly solving for $\rho$ from the master
equation, we determine analytical expressions of eigenvalues and
eigenvectors of $\rho^{T_A}$ (the partial transposition of density
matrix $\rho$). In Fock space, $\rho^{T_A}$ is shown to maintain a
block diagonal structure as the system evolves. In addition, we
discover that the decoherence induced by the baths would break the
degeneracy of $\rho^{T_A}$, and leads to a novel set of eigenvectors
for the construction of entanglement witness operators. Such
eigenvectors are shown to be time-independent, which is a signature
of robust entanglement of two-mode squeezed states in the presence
of noise.

\end{abstract}


\maketitle
\section{Introduction}

Optical two-mode squeezed vacuum (TMSV) has been a major source of
continuous-variable entanglement for quantum communication
\cite{braunstein}. In recent years, intriguing applications such as
quantum teleportation \cite{teleport1, teleport2, teleport3,
teleport4, teleport5} and quantum dense coding \cite{dense1, dense2}
have been demonstrated experimentally with TMSV. Theoretically, it
is also known that TMSV maximizes the EPR correlation when a fixed
amount of entanglement is given \cite{cirac2}. In order to exploit
fully the non-classical properties of such entangled light fields,
it is important to understand decoherence effects as they propagate
through noisy environments \cite{hiroshima,positionspace1, dodd,
welsch, Enk, wigner0, ban, wigner1, wigner2, char1}. This belongs to
a more subtle topic involving the characterization and
quantification of mixed state entanglement in general.

For bipartite systems, Peres and Horodecki have developed a
powerful criterion of entanglement, which is known as the PPT
(positive partial transposition) criterion \cite{PPTperes,
PPThoro1, PPThoro}. If the partial transposition of a density
matrix (denoted by $\rho^{T_A}$) has one or more negative
eigenvalues, then the state is an entangled state.  Physically,
the partial transposition for separable states can be considered
as a time-reversal operation, and one can construct a variety of
uncertainty relations serving as indicators of entanglement
\cite{Agarwal,vogel}. For two-mode Gaussian states such as TMSV,
PPT provides a necessary and sufficient condition of separability
\cite{Simon,LMDuan}.

The dynamics of disentanglement of TMSV in various noisy situations
has been addressed by several authors recently
\cite{hiroshima,positionspace1, dodd, welsch, Enk, wigner0, ban,
wigner1, wigner2, char1}. The fact that an amplitude damped TMSV
remains Gaussian enables an elegant description of entanglement
based on the properties of covariance matrix associated with the
density operators \cite{parisbook}. In particular, from the
time-dependent solution of Wigner function \cite{wigner0, ban,
wigner1} or the corresponding characteristic function \cite{char1},
one can quantify the degradation of entanglement by calculating the
negativity \cite{monotone} and relative entropy \cite{entropy}. It
is now known that for an initial TMSV at a non-zero thermal bath,
quantum entanglement vanishes completely in a finite time
\cite{parisbook}.

However, we notice that there are much less investigations
addressing the structure of $\rho^{T_A}$ directly, and yet
$\rho^{T_A}$ is what the PPT criterion originally based upon. Since
$\rho^{T_A}$ could manifest differently in various basis, the study
of $\rho^{T_A}$ in Fock space, for example, could reveal
entanglement properties not easily found by the Wigner function
method \cite{detect}. An example we notice is the construction of
entanglement witness operators via the projectors formed by the
eigenvectors of $\rho^{T_A}$ with negative eigenvalues
\cite{entneg}. Such entanglement witness operators, which correspond
a variety of observables for the detection of entanglement, are
determined by $\rho^{T_A}$.

The main purpose of this paper to indicate some key features of
decohered entanglement as revealed by eigenvalues and eigenvectors
of $\rho^{T_A}$. Our analysis will concentrate on the structures of
eigenvectors in Fock space, which is also where interesting
non-local correlations of continuous-variable systems can be
observed \cite{bellpseudospin}. For a TMSV under the influence of
amplitude damping (or gaining in an amplifier), we solve for the
time evolution of $\rho$ and determine the exact eigenvectors and
eigenvalues of $\rho^{T_A}$ analytically. These eigenvectors are
shown to have a strong correlation in photon numbers, and hence
$\rho^{T_A}$ is a block diagonal matrix in Fock space. Therefore
witness operators associated with each block involve only a finite
number of Fock vectors, which implies that the detection of
entanglement can only require a small portion of the Hilbert space.
This is in contrast to entanglement detection based on uncertainty
relations in which the entire Hilbert space is usually involved
\cite{Agarwal,vogel,Simon,LMDuan}. In this sense the eigenvectors of
$\rho^{T_A}$ access the entanglement signatures `locally', which is
a complement to `global' characterization (of Gaussian states) using
covariance matrices. As we shall see below, as long as the initial
state is a TMSV, the corresponding eigenvectors do not change with
time, indicating that the entanglement carried by TMSV is robust
against amplitude damping.

\section{Master equation and solution}

To begin with, we consider the time evolution of an initial TMSV,
each coupled with a separate phase-insensitive linear bath. In terms
of the annihilation operators $a$ and $b$ of the two modes, the
master equation governing the dynamical process is \cite{mandel}:
\begin{eqnarray} \label{eq:MASTER}
\dot{\rho} &=& G(2 a^\dag \rho a - aa^\dag \rho -\rho aa^\dag + 2
b^\dag \rho b - bb^\dag \rho -\rho bb^\dag)
\nonumber \\
&+&  L (2 a \rho a^\dag - a^\dag a \rho -\rho a^\dag a + 2 b \rho
b^\dag - b^\dag b \rho -\rho b^\dag b), \nonumber \\
\end{eqnarray}
where $G$ and $L$ are the gain and loss parameters respectively,
both having a dimension of $\mbox{time}^{-1}$. Depending on the
values of $G$ and $L$, the master equation describes amplifying or
damping effects due to the coupling with the baths. For dissipation
in thermal baths, each of temperature $T$, we have the parameters
$G=\frac{\gamma}{2} n_{th}$ and $L=\frac{\gamma}{2} (n_{th}+1)$,
where $n_{th}=\frac{1}{\exp({h\omega/kT})-1}$ is the average number
of photons in each of the modes (with frequency $\omega$) at thermal
equilibrium, and $\gamma/2$ is the decay rate of the mode
amplitudes. In this paper we focus on the initial TMSV with the
squeezing parameter $r> 0$:
\begin{eqnarray}
    \left|\psi(0)\right\rangle=\exp[r(a^\dag b^\dag-ab)]
    \left|00\right\rangle= \sqrt{1-\lambda^2} \sum^{\infty}_{n=0}
    \lambda ^n \left|nn\right\rangle,
\end{eqnarray}
where $\lambda\equiv\tanh r$ and $\left|00\right\rangle$ is the
two-mode vacuum state.

\subsection{Block structures of $\rho^{T_A}$ in Fock space}

\begin{figure}
\includegraphics [width=7 cm] {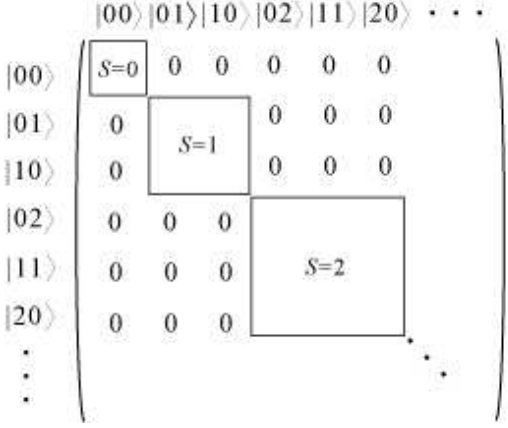}
\caption{\label{fig:submatrix} The sub-matrix structure of
$\rho^{T_A}$, with initial TMSV, in the Fock basis.}
\end{figure}

To investigate the entanglement properties of the density matrix
$\rho$, we study its partial transposition $\rho^{T_A}$. The
$\rho^{T_A}$ is of infinite dimension, however, by examining the
master equation in the Fock basis, block structures of
$\rho^{T_A}$ can be identified. Let us denote the matrix elements
of $\rho^{T_A}$ by
\begin{equation}
{\rho^{T_A}}_{n,m,p,q} = \left\langle nm| \rho^{T_A}
|pq\right\rangle = \left\langle pm| \rho |nq\right\rangle,
\end{equation}
which is governed by the following differential equation:
\begin{eqnarray} \label{eq:MASTEReltTA}
{\dot{\rho}^{T_A}}_{n,m,p,q} &=&G[2\sqrt{np}
{\rho^{T_A}}_{n-1,m,p-1,q} \nonumber \\ && \ \ + 2
\sqrt{mq}{\rho^{T_A}}_{n,m-1,p,q-1}
\nonumber \\
&& \ \ -
(n+m+p+q+4){\rho^{T_A}}_{n,m,p,q}]\nonumber \\
&& + L[2\sqrt{(n+1)(p+1)} {\rho^{T_A}}_{n+1,m,p+1,q} \nonumber \\
&& \ \ + 2 \sqrt{(m+1)(q+1)}{\rho^{T_A}}_{n,m+1,p,q+1}\nonumber
\\&& \ \ - (n+m+p+q){\rho^{T_A}}_{n,m,p,q}].
\end{eqnarray}
and
\begin{eqnarray}
{\rho^{T_A}}_{n,m,p,q}(t=0) = \delta_{pm}\delta_{nq}(1-\lambda^2)
\lambda^{m+n}.
\end{eqnarray}
corresponds to the initial condition (2).

It can be seen from Eq.~(\ref{eq:MASTEReltTA}) that each element
${\rho^{T_A}}_{n,m,p,q}(t)$ is coupled with elements
${\rho^{T_A}}_{n + l,m + k,p + l,q + k}(0)$ only, for integers $l$
and $k$. Therefore ${\rho ^{T_A}} _{n + l,m + k,m + l,n + k} \left(
t \right)$ are the only non-zero elements at any time $t>0$ because
of the initial condition. By noting that the sum of the first two
indices equal to that of the last two, we can group all non-zero
elements ${\rho ^{T_A}} _{n + l,m + k,m + l,n + k} \left( t \right)$
into sub-matrices ${\cal M}_S$ according to the sum index $S=n + l +
m + k$, i.e., ${\rho ^{T_A}}_{n + l,m + k,m + l,n + k} \left( t
\right)$ is contained in ${\cal M}_{n + l + m + k}$. We can
therefore express $\rho^{T_A}$ in a direct sum of ${\cal M}_S$ as
follows:
\begin{equation}
\rho^{T_A}(t)=\bigoplus^{\infty}_{S=0} {\cal M}_S (t),
\end{equation}
where the sub-matrix ${\cal M}_S$ has a dimension of $S+1$, since
elements in ${\cal M}_S$ have its first two indices as $\{0,S\}$,
$\{1,S-1\}$, ... , $\{S,0\}$. Fig. \ref{fig:submatrix} shows the
sub-matrix structure of $\rho^{T_A}$. Note that characteristic sum
$S$ is equal to the total number of photons that the two modes
contain. From Eq.~(\ref{eq:MASTEReltTA}) we observe that
probabilistic flow occurs between elements in neighboring
sub-matrices, with emission or absorption of one photon in one of
the modes at one time.

\begin{figure}
\includegraphics [width=8 cm] {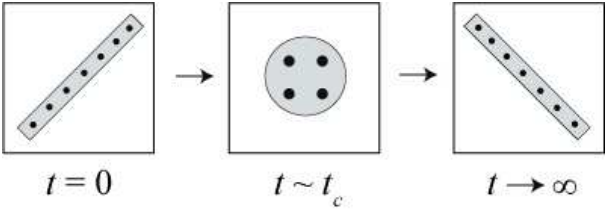}
\caption{\label{fig:evol} (Color online) Schematic diagram showing
the evolution of the distribution of elements in a sub-matrix of
$\rho^{T_A}$, assuming the baths are thermal baths. The $t_c$ is the
critical time for disentanglement. }
\end{figure}

The time evolution of a typical sub-matrix of $\rho^{T_A}$ is
illustrated schematically in Fig. \ref{fig:evol}, which will be
discussed in detail in the later part of the paper. Initially, only
opposite-diagonal elements are present, having the magnitude as
$\lambda^S$. As time increases, element flows from neighboring
sub-matrices, and disentanglement of the sub-matrices occurs at a
critical time $t=t_c$ (Section \ref{sect:eigen}). In the case of
thermal bath, $\rho^{T_A}$ evolves into a diagonal form in the long
time limit, settling as the thermal equilibrium state
${\rho^{T_A}}_{n,m,p,q}=\delta_{np}\delta_{mq}\frac{1}
{(n_{th}+1)^2}(\frac{n_{th}} {n_{th}+1})^{n+m}$.

\subsection{Analytic solution of $\rho$ in position space}

To analyze the properties of $\rho^{T_A}$, it is more convenient to
first determine $\rho$ in position space and then make the
transformation to Fock space. The position space method of finding
$\rho$ was previously employed in Ref.
\cite{positionspace1,positionspace2} in studying entanglement in
various oscillator systems. In this subsection, we present an
explicit solution of master equation (1) with an initial TMSV. We
remark that our method is different from that given in
\cite{positionspace1}, as the latter involves a Fourier transform of
the density matrix, i.e., the momentum space. Here we solve the
density matrix entirely in position space (Appendix A). This turns
out to be more convenient for the real symmetric Gaussian states
considered here, since fewer differential equations are involved. In
addition, the resultant solution is more transparent for further
analysis of eigenvectors in the next section.

Let us denote the `position' operators as $x =
\frac{1}{\sqrt{2}}(a+a^{\dag})$ and $y =
\frac{1}{\sqrt{2}}(b+b^{\dag})$, and define
\begin{equation}
\rho (x_1,y_1;x_2,y_2;t) \equiv \left\langle x_1,y_1 \right| \rho(t)
\left| x_2,y_2 \right\rangle,
\end{equation}
then the master equation (1) becomes,
\begin{eqnarray}\label{eq:posMASTER}
\dot \rho
&=&-\frac{1}{2}[L(x_1^2+x_2^2-2x_1x_2+y_1^2+y_2^2-2y_1y_2-4
\nonumber \\ && -\partial^2_{x_1}-\partial^2_{x_2}-\partial^2_{y_1}
-\partial^2_{y_2}-2\partial_{x_1}\partial_{x_2}
-2\partial_{y_1}\partial_{y_2} \nonumber
\\ && -2x_1\partial_{x_2}-
2x_2\partial_{x_1}-2y_1\partial_{y_2}-2y_2\partial_{y_1})\nonumber\\
&& +G(x_1^2+x_2^2-2x_1x_2+y_1^2+y_2^2-2y_1y_2+4 \nonumber \\
&& -\partial^2_{x_1}-
\partial^2_{x_2}-\partial^2_{y_1}-\partial^2_{y_2}-2\partial_{x_1}
\partial_{x_2} -2\partial_{y_1}\partial_{y_2} \nonumber \\
&& +2x_1\partial_{x_2}+2x_2\partial_{x_1}
+2y_1\partial_{y_2}+2y_2\partial_{y_1})] \rho.
\end{eqnarray}
For an initial state (2), $\rho (x_1,y_1;x_2,y_2;t)$ takes a
Gaussian form at any time $t$,
\begin{eqnarray}\label{eq:rho} \rho (x_1,y_1;x_2,y_2;t)&=
&\Xi(t) \exp [-A(t)(x_1^2+x_2^2+y_1^2+y_2^2) \nonumber \\
&& +B(t)(x_1 y_1+x_2 y_2) \nonumber
\\&& + C(t)(x_1 x_2+y_1 y_2) \nonumber \\
&& + D(t)(x_1 y_2+x_2 y_1)],
\end{eqnarray}
where $A(t)$, $B(t)$, $C(t)$ and $D(t)$ are real time-dependent
coefficients, and the normalization factor is:
\begin{eqnarray}
\Xi(t)=\frac{1}{\pi}\sqrt{[2A(t)-C(t)]^2-[B(t)+D(t)]^2}.
\end{eqnarray}
By substituting Eq.~(\ref{eq:rho}) into the master equation, the
coefficients are found to obey a set of coupled equations that can
be solved analytically (Appendix \ref{app1}). For the TMSV
considered here, we have,
\begin{eqnarray}\label{eq:squecoeff}
A(t)&=&\frac{1}{4}[2A_0\eta + \frac{G+L}{G-L} (\eta - 1) +
\frac{\left\langle  {x}^2\right\rangle_t}{2(\left\langle
 {x}^2\right\rangle_t^2-\left\langle
 {x} {y}\right\rangle_t^2)}],\nonumber\\
B(t)&=&\frac{1}{2}[B_0\eta+\frac{\left\langle  {x} {y}
\right\rangle_t}{2(\left\langle  {x}^2\right\rangle_t^2
-\left\langle  {x} {y}\right\rangle_t^2)}], \nonumber\\
C(t)&=&\frac{1}{2}[2A_0\eta + \frac{G+L}{G-L} (\eta - 1) -
\frac{\left\langle  {x}^2\right\rangle_t}{2(\left\langle
 {x}^2\right\rangle_t^2-\left\langle  {x} {y}
\right\rangle_t^2)}],\nonumber\\
D(t)&=&\frac{1}{2}[-B_0\eta+\frac{\left\langle
 {x} {y}\right\rangle_t}{2(\left\langle
 {x}^2\right\rangle_t^2-\left\langle
 {x} {y}\right\rangle_t^2)}].
\end{eqnarray}
Here $A_0= \frac{1}{2}\cosh 2r$, $B_0= \sinh 2r$ and $\eta (t)
=\exp[2(G-L)t]$ are defined, and the expectation values are given
by,
\begin{eqnarray}
\left\langle  {x}^2\right\rangle_t &=& A_0 \eta
+  \frac{G+L}{2(G-L)}(\eta - 1), \nonumber\\
\left\langle  {x} {y}\right\rangle_t &=& \frac{B_0}{2} \eta.
\end{eqnarray}

\section{Properties of $\rho^{T_A}$}

According to PPT criterion, the appearance of negative eigenvalues
of $\rho^{T_A}$ is a signature of entanglement. In this section, we
solve the eigenvectors and eigenvalues of $\rho^{T_A}$ as the system
evolves. Then we discuss how decoherence affects the entanglement
properties of $\rho^{T_A}$. The eigenvalues and eigenvectors of
$\rho^{T_A}$ are defined by:
\begin{eqnarray}\label{eq:eigensysdef}
&& \int\int\rho^{T_A}(x_1,y_1;x_2,y_2;t)\varphi_{n,m}(x_2,y_2;t)
dx_2dy_2 \nonumber \\ && = \xi_{n,m}(t)\varphi_{n,m}(x_1,y_1;t).
\end{eqnarray}
Our main technique of solving the eigen-problem is the use of Mehler
formula which expands a double Gaussian function into a series of
orthogonal functions. After some calculations (see Appendix
\ref{app2}), we obtain the expression of eigenvalues,
\begin{eqnarray}\label{eq:eigenvalsque}
\xi_{n,m}(t)= \Xi(t)\pi
\frac{(\sqrt{\alpha_1}-\sqrt{\beta_1})^n}{(\sqrt{\alpha_1}
+\sqrt{\beta_1})^{n+1}}\frac{(\sqrt{\alpha_2}-
\sqrt{\beta_2})^{m}}{(\sqrt{\alpha_2}+\sqrt{\beta_2})^{m+1}}
\end{eqnarray}
and the corresponding eigenvectors,
\begin{eqnarray}\label{eq:eigenvecsque1}
\varphi_{n,m}(x_1,y_1)&=&\frac{1}{\sqrt{2^{n+m}n!m! \pi}}
 H_n(\frac{x_1-y_1}{\sqrt{2}})
H_m(\frac{x_1+y_1}{\sqrt{2}})
\nonumber \\
&&\times
 \exp[-\frac{1}{2} (x_1^2+y_1^2)]
\end{eqnarray}
where $H_n$ are the Hermite polynomials. In
writing Eq.~(\ref{eq:eigenvalsque}), we have defined
\begin{eqnarray}
&& \alpha_1(t)=\frac{1}{4}[2A(t)-B(t)+C(t)+D(t)], \nonumber \\
&& \beta_1(t)=\frac{1}{4}[2A(t)+B(t)-C(t)+D(t)],\nonumber \\
&& \alpha_2(t)=\frac{1}{4}[2A(t)+B(t)+C(t)-D(t)],\nonumber \\
&& \beta_2(t)=\frac{1}{4}[2A(t)-B(t)-C(t)-D(t)]
\end{eqnarray}
and the solution of $A$, $B$, $C$ and $D$ are given by
Eq.~(\ref{eq:squecoeff}). Note that in writing
Eq.~(\ref{eq:eigenvecsque1}) from (B8), we have used the fact that
for TMSV, $\alpha_1\beta_1=\alpha_2\beta_2=\frac{1}{16}$  for all
time $t\geq0$.

We now transform the eigenvectors from the position space to the
Fock space. Note that $\rho^{T_A}$ is a basis dependent operation.
The eigenvectors of $\rho^{T_A}$ defined in two different basis sets
do not transform directly. An exception is the case when the two
sets of basis vectors transform by a real unitary matrix
\cite{PTbasisdep}, which is the case here. This allows us to write
down the eigenvector $\left|\varphi_{n,m}\right\rangle$ in Fock
space from Eq.~(\ref{eq:eigenvecsque1}):
\begin{equation}
\left|\varphi_{n,m}\right\rangle=[\frac{1}{\sqrt{n!}}
(\frac{a^\dag-b^\dag}{\sqrt{2}})^n ]
[\frac{1}{\sqrt{m!}}(\frac{a^\dag+b^\dag}{\sqrt{2}})^m]
\left|00\right\rangle,
\end{equation}
which can be connected to the sub-matrices of $\rho^{T_A}$ (Fig.
\ref{fig:submatrix}) via the photon number sum $S\equiv m+n$ to
label the eigenket, so that
\begin{eqnarray}\label{eq:eigenvecsque}
    \left|\varphi_{n,S-n}\right\rangle &=&\frac{1}{\sqrt{2^S}
    \sqrt{n!(S-n)!}} \nonumber \\
&& \times
    \sum^{S}_{j=0}\Gamma_{S,n,j}\sqrt{j!(S-j)!}
    \left|j,S-j\right\rangle.
\end{eqnarray}
Here we have used the abbreviation
\begin{equation}
\Gamma_{S,n,j} \equiv
\sum^{\min(j,n)}_{k=0}(-1)^{n-k}C^{S-n}_{j-k}C^{n}_{k},
\end{equation}
with $C^{S}_{r}\equiv\frac{S!}{r!(S-r)!}$. In this way
$\left|\varphi_{n,S-n}\right\rangle$ and $\xi_{n,S-n}$
$(n=0,1,...,S)$ are the $n^{th}$ eigenvector and eigenvalue of the
block with characteristic sum $S$.

\subsection{Evolution of negative eigenvalues}\label{sect:eigen}

\begin{figure}
\includegraphics [width=8 cm] {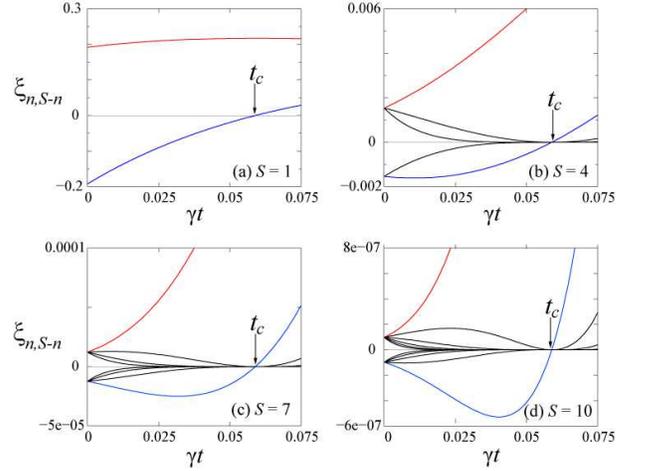}
\caption{\label{fig:negeigen} (Color online) The eigenvalues of
sub-matrices of $\rho^{T_A}$ of an amplifier system with $G=1.5
\gamma$, $L=0.5\gamma$ and initial squeezing factor $r=\tanh^{-1}
0.2$, with characteristic sum (a) $S=1$, (b) $S=4$, (c) $S=7$, and
(d) $S=10$. Red line indicates the $n=0$ eigenvalue which does not
turn zero at $t_c$, while blue line indicates the most negative
eigenvalue with $n=1$.}
\end{figure}

By inspecting Eq.~(\ref{eq:eigenvalsque}), we find that all the
negative eigenvalues in each block share the same value at $t=0$
(Fig. \ref{fig:negeigen}). The same is true also for positive
eigenvalues. However, such a strong degeneracy is broken by coupling
with the baths. This is illustrated in Fig. \ref{fig:negeigen} where
the time-dependence of individual eigenvalues in various block
indices $S$ is shown. Except at $t=0$ and at the critical time
$t_c$, we see that the negative eigenvalues possessing different
values.

It is important to observe that the eigenvalues are negative for odd
$n$, when $\sqrt{\alpha_1} < \sqrt{\beta_1}$, or in other words,
$B(t) > C(t)$. All eigenvalues turn zero at the same time, when we
have $B(t)=C(t)$, except for the only eigenvalue with $n=0$ in each
block. Such a critical time $t_c$ is given by,
\begin{equation} \label{eq:tc}
t_c  = \frac{{1}}{{2\left( {L-G} \right)}}\log \left( {\frac{{G +
L\lambda }}{{G\left( {1 + \lambda } \right)}}} \right)
\end{equation}
which is always positive finite as long as $G \not= 0$. For the case
$G\rightarrow 0$, we have $t_c \rightarrow \infty$. We remark that
the disentanglement time $t_c$ was previous obtained in Ref.
\cite{LMDuan} for thermal baths, here we obtained a general
expression (\ref{eq:tc}) that applies to linear amplifiers as well.
In particular, in the case when gain and loss parameters are equal,
i.e., $G=L$, the critical time can be reduced to $t_c =
\frac{1}{2G}\frac{\lambda}{1+\lambda} > 0$.

We point out that at the time of disentanglement $t=t_c$, there is
only {\em one} non-zero eigenvalue (with the index $n=0$) in each
sub-matrix (Fig. \ref{fig:negeigen}). Therefore $\rho^{T_A}$ at the
critical time is highly degenerate, and the corresponding symmetry
property of $\rho^{T_A} (t=t_c)$ is indicated in the relation:
\begin{eqnarray}
&& {\rho^{T_A}}_{j,S-j,j,S-j}  ={\rho^{T_A}}_{j,S-j,S-j,j} \nonumber \\
&& ={\rho^{T_A}}_{S-j,j,j,S-j}={\rho^{T_A}}_{S-j,j,S-j,j}.
\end{eqnarray}
This results in the symmetric distribution of elements as shown in
Fig. \ref{fig:evol} schematically.

As a further remark, it is interesting that negative eigenvalues may
not necessarily be monotones over time. This can be seen by
differentiating Eq.~(\ref{eq:eigenvalsque}) and looking at the
initial rate:
\newcommand\sech{\mathop{\mathrm{sech}}}
\newcommand\csch{\mathop{\mathrm{csch}}}
\begin{eqnarray} \label{eq:eigendiff}
\left. \dot{\xi}_{n,S-n} \right|_{t = 0} &=&(-1)^n \tanh^S r \csch r
{\sech} ^3 r
\nonumber\\ &&  \times \{(G-L)(S-2n) \nonumber \\
&& \ \  + (G+L)(S-2n) \cosh 2r \nonumber
\\ &&\ \   -[LS+G(2+S)]\sinh 2r\},
\end{eqnarray}
which can result in a negative value for certain parameters, i.e.,
some eigenvalues of odd $n$ can become more negative over time (Fig.
\ref{fig:negeigen}d). An exceptional case is when $G=0$, where we
find that for odd $n$, the derivative at $t=0$ must be positive by inspecting
Eq.~(\ref{eq:eigendiff}).

\begin{figure}
\includegraphics [width=7 cm] {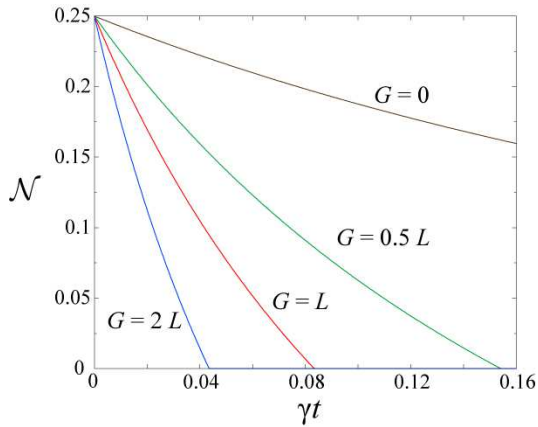}
\caption{\label{fig:neg} (Color online) The negativity ${\cal N}$ of
$\rho^{T_A}$ with different $G$, fixing the parameters $L= \gamma$
and initial squeezing factor $r=\tanh^{-1} 0.2$.}
\end{figure}

\subsection{Negativity and sub-negativity}

Negativity ${\cal N}$ serves as a computable measure of entanglement
defined by the trace norm of $\rho^{T_A}$ minus 1 divided by 2
\cite{monotone}. For separable states, $\rho^{T_A}$ is still a
density matrix with trace 1 and hence ${\cal N}=0$. However, for
non-separable states with negative $\rho^{T_A}$, we have ${\cal N}
>0 $. Specifically, ${\cal N}$ equals the sum of the
absolute value of negative eigenvalues of $\rho^{T_A}$
\cite{monotone,negeigenadd}. For the system considered in this
paper, ${\cal N}$ reads,
\begin{eqnarray}
{\cal N} =\frac{\pi\Xi(t)}{8}(\frac{1}{\sqrt{\alpha_1 \beta_2}}
-\frac{1}{\sqrt{\beta_1 \beta_2}}).
\end{eqnarray}
Alternatively, ${\cal N}$ can be derived from the symplectic
spectrum of the covariance matrix associated with the density
operator \cite{parisbook}. In Fig. \ref{fig:neg} we show the time
evolution of negativity ${\cal N}$ for initial TMSV with different
$G$ parameters. An example for the $G=0$ case is the
zero-temperature bath dissipation scenario. Fixing $L$, we observe
that a larger gain $G$ leads to a shorter $t_c$.

\begin{figure}
\includegraphics [width=8 cm] {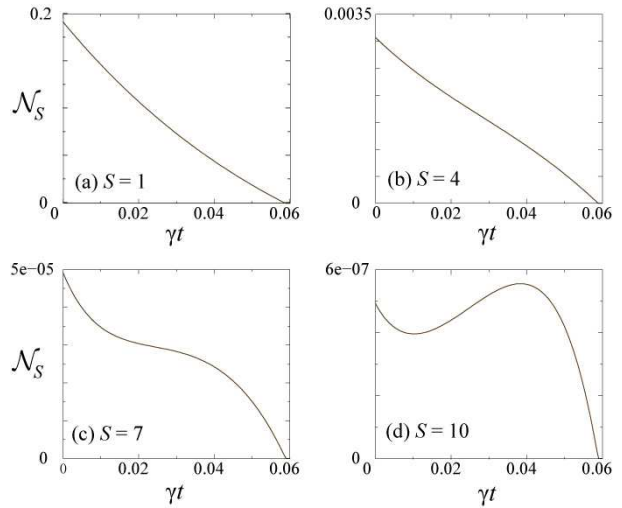}
\caption{\label{fig:blockneg} The negativity of sub-matrices ${\cal
N}_S$ of $\rho^{T_A}$, for an amplifier system with $G=1.5 \gamma$,
$L=0.5 \gamma$ and initial squeezing factor $r=\tanh^{-1} 0.2$, with
characteristic sum (a) $S=1$, (b) $S=4$, (c) $S=7$, and (d) $S=10$.}
\end{figure}

We can also calculate the negativity in a sub-matrix ${\cal M}_S$,
which measures the contribution of entanglement from the
corresponding block that builds up $\rho^{T_A}$. Specifically, the
sub-matrix negativity ${\cal N}_S$ is defined the same way as
negativity ${\cal N}$ but restricted to the sub-matrix of
$\rho^{T_A}$ with the characteristic sum $S$. To our knowledge, such
a sub-matrix negativity, which requires the calculations of
individual eigenvalues, has not been discussed before. Explicitly,
${\cal N}_S$ takes the form:
\begin{equation}
{\cal N}_S = -\pi \Xi(t)\sum^{P}_{n=0}
\frac{(\sqrt{\alpha_1}-\sqrt{\beta_1})^{2n+1}}{(\sqrt{\alpha_1}
+\sqrt{\beta_1})^{2n+2}}\frac{(\sqrt{\alpha_2}-
\sqrt{\beta_2})^{S-2n-1}}{(\sqrt{\alpha_2}+\sqrt{\beta_2})^{S-2n}}
\end{equation}
where $P$ is defined as the integral part of $\frac{S-1}{2}$. In
Fig. \ref{fig:blockneg} the behavior of ${\cal N}_S$ of some blocks
is illustrated. It is surprising that that for higher sub-matrices
the corresponding negativity ${\cal N}_S$ may {\em increase} over
time. Physically, the increase of ${\cal N}_S$ is due to the
probability flow in Fock space arising from damping or amplifying
mechanisms. Since each sub-matrix of $\rho^{T_A}$ does not
necessarily conserve probability (i.e., the trace of ${\cal M}_S$ is
not a constant), it is possible that some blocks could have their
negativity increasing with time.

However, the increase of ${\cal N}_S$ does not violate the fact that
the overall negativity ${\cal N}$ of $\rho^{T_A}$ is an entanglement
monotone that does not increase under LOCC (the master equation
corresponds to local operations). Eq.~(\ref{eq:eigenvalsque})
reveals that eigenvalues of higher blocks are of smaller order of
magnitude. As we see in Fig. \ref{fig:blockneg}, although negativity
of individual higher sub-matrices may increase over time, their
contribution for negativity is smaller by several orders than the
sub-matrices with lower $S$, and therefore the overall negativity is
still monotonic decreasing.

\subsection{Robust structure of entanglement witness}

\begin{figure}
\includegraphics [width=6 cm] {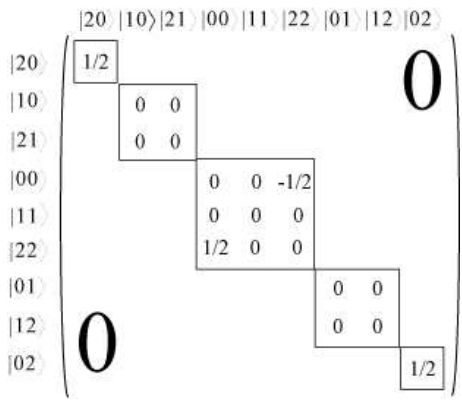}
\caption{\label{fig:witness1} Example of entanglement witness ${\cal
W}_{2,1}$.}
\end{figure}

\begin{figure}
\includegraphics [width=6 cm] {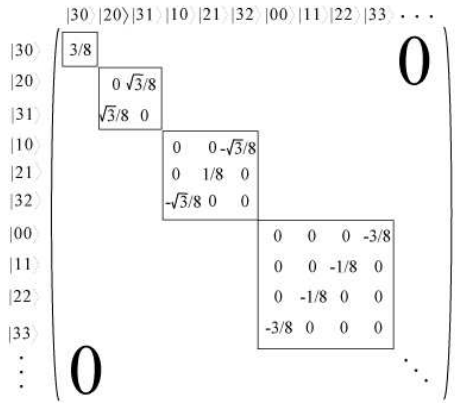}
\caption{\label{fig:witness2} Part of the entanglement witness
${\cal W}_{3,1}$, showing the blocks with characteristic difference
from -3 to 0. ${\cal W}_{3,1}$ is symmetric about the top-right to
bottom-left diagonal.}
\end{figure}

An entanglement witness operator ${\cal W}$ is designed for the
detection of entanglement such that $Tr(\rho {\cal W})<0$ for some
non-separable states $\rho$, but $Tr(\rho_{sep} {\cal W}) \ge 0$ for
all separable states $\rho_{sep}$ \cite{PPThoro1}. For each
eigenvector $|\phi \rangle$ of $\rho^{T_A}$ with a negative
eigenvalue, one can construct an ${\cal W}$ by ${\cal W}= |\phi
\rangle \langle \phi |^{T_A}$ \cite{entneg} meeting the criteria
above. In our system, we can construct a family of ${\cal W}$ from
eigenvectors $\left|\varphi_{n,S-n}\right\rangle$ with odd $n$
accordingly, i.e., ${\cal W}_{S,n} =
\left|\varphi_{n,S-n}\right\rangle\left\langle
\varphi_{n,S-n}\right|^{T_A}$. From Eq.~(\ref{eq:eigenvecsque}), the
explicit form of ${\cal W}_{S,n}$ reads,
\begin{eqnarray}\label{eq:witnesssque}
    {\cal W}_{S,n} &=&\sum^{S}_{j=0}\sum^{S}_{l=0}\{\frac{1}{{2}^S
    {n!(S-n)!}}\Gamma_{S,n,j}\Gamma_{S,n,l}\nonumber \\
    &&\times\sqrt{j!(S-j)!l!(S-l)!}\}
    \left|l,S-j\right\rangle\left\langle
    j,S-l\right|, \nonumber \\
\end{eqnarray}
which shows that ${\cal W}_{S,n}$ operates in only a finite
dimension in the Fock space. Note that $Tr(\rho {\cal W}_{S,n})=
\xi_{n,S-n}$, and $\xi_{1,S-1}$ is the most negative eigenvalue in
each sub-matrix, therefore ${\cal W}_{S,1}$ provides the most
significant entanglement detection among all witnesses constructed
from vectors lying within the sub-matrix ${\cal M}_S$. Some examples
for entanglement witnesses with $n=1$ are shown in Fig.
\ref{fig:witness1} and Fig. \ref{fig:witness2}.

We observe that ${\cal W}$ also has a block diagonal structure in
which non-zero elements are: $\left\langle j,S-l\right|{\cal
W}_{S,n}\left|l,S-j \right\rangle$. This allows us to divide ${\cal
W}$ into sub-blocks $\Lambda_K$, with each sub-block characterized
by a difference $K\in[-S,S]$:
\begin{equation}
 {\cal W}_{S,n}=\bigoplus^{S}_{K=-S} \Lambda_K,
\end{equation}
where the element $\left\langle j,S-l\right|{\cal
W}_{S,n}\left|l,S-j \right\rangle$ lies in the sub-block with
$K=S-j-l$. The explicit form of $\Lambda_K$ is
\begin{eqnarray}
\Lambda_K&=&\sum^{\min(S,S-K)}_{l=\max(0,-K)}\{\frac{1}{{2}^S
    {n!(S-n)!}}\Gamma_{S,n,S-K-l}\Gamma_{S,n,l}\nonumber
    \\ &&\times\sqrt{(S-K-l)!(K+l)!l!(S-l)!}\} \nonumber \\
    && \ \ \ \left|l,K+l\right\rangle\left\langle
    S-K-l,S-l\right|,
\end{eqnarray}
having a dimension of $S-|K|+1$. We remark that the simplest
$2\times 2$ entanglement witness ${\cal W}_{1,1}$ was constructed in
\cite{witnessexplicit} using a different approach. Here our general
${\cal W}_{S,n}$ applies to all $S$ and $n$.


Finally, let us emphasize the robustness feature of TMSV against
decoherence. We have seen from Eq.~(\ref{eq:eigenvecsque})
explicitly that the eigenvectors
$\left|\varphi_{n,S-n}\right\rangle$ remain unchanged with time. The
time-independent $\left|\varphi_{n,S-n}\right\rangle$ suggests that
TMSV is robust against noise, in the sense that structure of
entanglement witness ${\cal W}_{S,n}$ is preserved. The degradation
of entanglement would only affect the eigenvalues. We stress that
such a time-independent property of eigenvectors is specific to
initial TMSV, and does not hold for arbitrary initial states in
general. In Appendix \ref{app2}, we derive the eigenvectors of
$\rho^{T_A}$ evolving from an initial two-mode symmetric Gaussian
states with {\em arbitrary} real coefficients $A_0$, $B_0$, $C_0$
and $D_0$. We find that the time dependence of the eigenvectors
arises solely from the evolution of the factors $\alpha_1\beta_1$
and $\alpha_2\beta_2$. Because of the system-bath interactions,
these factors are generally dependent on time. However, the state
evolving from TMSV is an exception in which one can show that the
corresponding $\alpha_1\beta_1$ and $\alpha_2\beta_2$ equal the
constant $1/16$ at all times.

\section{Conclusion}

To summarize, by deriving an exact analytic solution for
$\rho^{T_A}$ and examining its eigenvectors, we discover several
important features about the loss of entanglement of a TMSV suffered
from decoherence. Both amplitude damping and amplification effects
have been included in our analysis. Throughout the decoherence
process, the block diagonal structure of $\rho^{T_A}$ is shown to be
maintained in Fock basis. As each block spans only a finite portion
of the Fock space, the existence of negative eigenvalues in a block
implies that entanglement information can `survive' in the
corresponding photon-number subspace. If quantum entanglement of the
system is to be destroyed completely, then all the blocks have to be
made positive. In other words, by simply mixing the system with
another state involving finite Fock space would not destroy the
entanglement. For the decoherence process considered in this paper,
all the blocks turn positive at a critical time $t_c$, which agree
with the previous analysis based on the covariance matrix. At $t <
t_c$, we derived an explicit expression of the negativity as well as
the negativity of sub-matrices in order to characterize the
time-dependence of entanglement. Interestingly, the negativity of
some sub-matrices could increase with time when $G$ is non-zero,
although the effect is weak according to our calculations. The
signature of entanglement in photon number subspace can be detected
by the witness operators ${\cal W}_{S,n}$, and we have constructed
${\cal W}_{S,n}$ explicitly in this paper. These witness operators
are also block diagonal, and most remarkably, they are
time-independent even though the system is under the influence of
noise. We interpret such a property as a kind of robust entanglement
structure inherited in TMSV.


\begin{acknowledgments}
This work is supported by the Research Grants Council of the Hong
Kong SAR, China (Project No. 401305).
\end{acknowledgments}

\appendix
\section{Time evolution of general real symmetric two-mode Gaussian
density operator}

\label{app1} In this Appendix, we consider the time evolution of
real symmetric two-mode Gaussian states, each coupling linearly with
a separate bath with general gain and loss parameters. The density
matrix is represented in position space, obeying the master equation
as in Eq.~(\ref{eq:posMASTER}). For the solution of the two-mode
real symmetric Gaussian state presented in Eq.~(\ref{eq:rho}), a
direct substitution leads to the following coupled differential
equations:
\begin{eqnarray} \label{eq:A}
\dot{A} &=&(G-L)C +\frac{1}{2}(G+L) [1-(2A-C)^2
-(B +D)^2 ], \nonumber\\
\dot{B}&=&-2(G-L)D-2(G+L)(2A-C)(B+D),
\nonumber\\
\dot{C}&=&4(G-L)A+(G+L) [ 1+(2A-C)^2+(B+D)^2 ],
\nonumber\\
\dot{D}&=&-2(G-L)B-2(G+L)(2A-C)(B+D).
\end{eqnarray}
Without loss of generality, we consider the  $G\not= L$ case. From
Eqs. (A1), $B-D$ and $2A+C$ have the simple solution,
\begin{eqnarray}
B(t)-D(t)&=&(B_0-D_0)\eta(t), \nonumber\\
2A(t)+C(t)&=&(2A_0+C_0)\eta(t) + \frac{G+L}{G-L} [\eta(t)-1], \nonumber \\
\end{eqnarray}
where $\eta (t) \equiv\exp[2(G-L)t]$ and the zero subscripts denote
the values at $t=0$. The other two combinations, $B+D$ and $2A-C$,
can be found by noting that they are related to the second moments
$\langle x^2\rangle $ and $\langle xy \rangle$  by:
\begin{eqnarray}
2A(t)-C(t)&=&\frac{\left\langle {x}^2\right\rangle_t}
{2[\left\langle {x}^2\right\rangle_t^2-\left\langle {x}{y}
\right\rangle_t^2]}, \nonumber\\
B(t)+D(t)&=&\frac{\left\langle {x}{y}\right\rangle_t}{2[\left\langle
{x}^2\right\rangle_t^2-\left\langle {x}{y}\right\rangle_t^2]}.
\end{eqnarray}
with the subscript $t$ denoting the value at time $t$. Under the
condition:
\begin{equation}
\left\langle a^2\right\rangle_t = \left\langle b^2\right\rangle_t =
\left\langle a^\dag b\right\rangle_t = 0
\end{equation}
which applies to TMSV, and from the Heisenberg equations of motions
of $a$ and $b$, the time-dependence of the second moments are given
by:
\begin{eqnarray}\label{eq:expx2}
\left\langle {x}^2\right\rangle_t &=& \left\langle
{x}^2\right\rangle_0
\eta(t) + \frac{G+L}{2(G-L)}[\eta(t) - 1],\nonumber\\
\left\langle {x} {y}\right\rangle_t &=& \left\langle
 {x} {y}\right\rangle_0 \eta(t).
\end{eqnarray}
Here the initial second moments are given by:
\begin{eqnarray}\label{eq:expx2ini}
\left\langle  {x}^2\right\rangle_0 &=& \frac{2A_0-C_0}
{2[(2A_0-C_0)^2-(B_0+D_0)^2]} ,\nonumber\\
\left\langle  {x} {y}\right\rangle_0 &=&
\frac{B_0+D_0}{2[(2A_0-C_0)^2-(B_0+D_0)^2]}.
\end{eqnarray}
Thus the time evolution of the coefficients of the two-mode
Gaussian state solution are as follows:
\begin{eqnarray}
A(t)&=&\frac{1}{4}[(2A_0+C_0)\eta + \frac{G+L}{G-L} (\eta - 1)
\nonumber \\ && + \frac{\left\langle
{x}^2\right\rangle_t}{2(\left\langle
{x}^2\right\rangle_t^2-\left\langle  {x} {y}\right\rangle_t^2)}],
\nonumber\\
B(t)&=&\frac{1}{2}[(B_0-D_0)\eta+\frac{\left\langle  {x} {y}
\right\rangle_t}{2(\left\langle  {x}^2\right\rangle_t^2-
\left\langle  {x} {y}\right\rangle_t^2)}], \nonumber\\
C(t)&=&\frac{1}{2}[(2A_0+C_0)\eta + \frac{G+L}{G-L} (\eta - 1)
\nonumber \\ && - \frac{\left\langle
{x}^2\right\rangle_t}{2(\left\langle {x}^2
\right\rangle_t^2-\left\langle  {x} {y}\right\rangle_t^2)}],
\nonumber\\
D(t)&=&\frac{1}{2}[-(B_0-D_0)\eta+\frac{\left\langle
 {x} {y}\right\rangle_t}{2(\left\langle
 {x}^2\right\rangle_t^2-\left\langle
 {x} {y}\right\rangle_t^2)}],
\end{eqnarray}
where the expectation values are given in Eq.~(\ref{eq:expx2}).

In the case of TMSV, the solution is reduced to Eq. (12) as the
initial coefficients satisfy:
\begin{eqnarray}
&& C_0=D_0=0 \\
&& 4A_0^2-B_0^2=1.
\end{eqnarray}
In particular, by noting that $A_0=\left\langle {x}^2\right\rangle_0
= \frac{1}{2} + \left\langle a^\dag a\right\rangle_0= \frac{1}{2} +
\frac{\lambda ^2}{1-\lambda ^2}$, we have $A_0= \frac{1}{2}\cosh
2r$.

\section{Derivation of eigenvectors and eigenvalues of $\rho^{T_A}$}\label{app2}
In position space, $\rho^{T_A}(x_1,y_1;x_2,y_2;t) =\rho
(x_2,y_1;x_1,y_2;t)$. The eigenvectors $\varphi_{nm}$ and
eigenvalues $\xi_{nm}$ of $\rho^{T_A}$ are defined by
Eq.~(\ref{eq:eigensysdef}). Applying the transformation
$x_j=(u_j+v_j)/{\sqrt{2}}$ and $y_j=(-u_j+v_j)/{\sqrt{2}}$
$(j=1,2)$, $\rho^{T_A}$ becomes a neat product of two double
Gaussians as follows:
\begin{eqnarray}\label{eq:rhonewbasis}
&& \rho^{T_A} (u_1,v_1;u_2,v_2;t) \nonumber \\
&& =\Xi(t) \exp [-\alpha_1(t) (u_1-u_2)^2 - \beta_1(t)
(u_1+u_2)^2]\nonumber \\& &\times\exp [-\alpha_2(t) (v_1-v_2)^2 -
\beta_2(t) (v_1+v_2)^2],
\end{eqnarray}
where $\alpha_1=\frac{1}{4}(2A-B+C+D)$, $\beta_1
=\frac{1}{4}(2A+B-C+D)$, $\alpha_2=\frac{1}{4}(2A+B+C-D)$ and
$\beta_2=\frac{1}{4}(2A-B-C-D)$. We apply Mehler's Formula twice,
one for the $u_1$, $u_2$ double Gaussian, and one for the $v_1$,
$v_2$ double Gaussian function. This would lead to the Schmidt
decomposition on $\rho^{T_A}$:
\begin{eqnarray}\label{eq:rhoschmidt}
&& \rho^{T_A} (u_1,v_1;u_2,v_2;t) = \Xi(t)
\frac{\pi}{4}(\frac{1}{\alpha_1\beta_1\alpha_2\beta_2})^\frac{1}{4}
\nonumber \\&&\times\sum^{\infty}_{n=0} \lambda_n f_n(u_1)f_n(u_2)
\sum^{\infty}_{m=0} \tilde \lambda_m \tilde f_m(v_1) \tilde
f_m(v_2),
\end{eqnarray}
where the Schmidt modes $f_n(u)$ and $\tilde f_m(u)$ are:
\begin{eqnarray}
f_n(u)&=& \frac{1}{\sqrt{2^{n-1}n!}}
(\frac{\sqrt{\alpha_1\beta_1}}{\pi})^\frac{1}{4} H_n
[2(\alpha_1\beta_1)^\frac{1}{4} u] \nonumber \\
&& \ \ \ \exp ({-2\sqrt{\alpha_1\beta_1} u^2}),\nonumber
\\ \tilde f_m(u) &=& \frac{1}{\sqrt{2^{m-1}m!}}
(\frac{\sqrt{\alpha_2\beta_2}}{\pi})^\frac{1}{4} H_m[2
(\alpha_2\beta_2)^\frac{1}{4} u] \nonumber \\
&& \exp ({-2\sqrt{\alpha_2\beta_2} u^2}),
\end{eqnarray}
and the coefficients $\lambda_n(t)$ and $\tilde \lambda_m(t)$
\begin{eqnarray}
 &&\lambda_n(t)=2(\alpha_1\beta_1)^\frac{1}{4}
\frac{(\sqrt{\alpha_1}-\sqrt{\beta_1})^n}{(\sqrt{\alpha_1}
+\sqrt{\beta_1})^{n+1}},\nonumber
\\&& \tilde \lambda_m(t)=2(\alpha_2\beta_2)^\frac{1}{4}\frac{(\sqrt{\alpha_2}
-\sqrt{\beta_2})^m}{(\sqrt{\alpha_2}+\sqrt{\beta_2})^{m+1}},
\end{eqnarray}
where $H_n(u)$ are the Hermite polynomials. Rearranging terms,
Eq.~(\ref{eq:rhoschmidt}) gives:
\begin{eqnarray}
\rho^{T_A} (x_1,y_1;x_2,y_2;t) \equiv &&
\sum_{n,m}\xi_{n,m}(t)\varphi_{n,m}(x_1,y_1;t) \nonumber
\\ && \ \ \times \varphi_{n,m}(x_2,y_2;t),
\end{eqnarray}
where the eigenvectors $\varphi_{n,m}$ are
\begin{eqnarray}\label{eq:eigenvec}
&& \varphi_{n,m}(x_1,y_1;t) \nonumber \\
&& =\frac{1}{\sqrt{2^{n-1}2^{m-1}n!m!}}
(\frac{\sqrt{\alpha_1\beta_1\alpha_2\beta_2}}{\pi^2})^\frac{1}{4}
\nonumber \\
&&\times H_n[2(\alpha_1\beta_1)^\frac{1}{4}
(\frac{x_1-y_1}{\sqrt{2}})]H_m[2(\alpha_2\beta_2)^\frac{1}{4}
(\frac{x_1+y_1}{\sqrt{2}})] \nonumber \\
&&\times \exp[-\sqrt{\alpha_1\beta_1} (x_1-y_1)^2]
\exp[-\sqrt{\alpha_2\beta_2} (x_1+y_1)^2], \nonumber \\
\end{eqnarray}
and the eigenvalues $\xi_{n,m}$ of $\rho^{T_A}$ are,
\begin{eqnarray}\label{eq:eigenval}
\xi_{n,m}(t)= \Xi(t)\pi
\frac{(\sqrt{\alpha_1}-\sqrt{\beta_1})^n}{(\sqrt{\alpha_1}
+\sqrt{\beta_1})^{n+1}}\frac{(\sqrt{\alpha_2}-\sqrt{\beta_2})^m}
{(\sqrt{\alpha_2}+\sqrt{\beta_2})^{m+1}}. \nonumber \\
\end{eqnarray}
These expressions of eigenvectors and eigenvalues are for any
time-dependent coefficients $A(t)$, $B(t)$, $C(t)$ and $D(t)$, i.e.,
applicable to states evolving from arbitrary initial values $A_0$,
$B_0$, $C_0$ and $D_0$. The special case with initial TMSV is given
in Eq.~(\ref{eq:eigenvalsque}) and Eq.~(\ref{eq:eigenvecsque}).


\begin{references}
\bibitem{PTbasisdep} Let $\left|\phi\right\rangle
\equiv\sum_{nm}c_{nm} |n\rangle |m \rangle$ be an eigenvector of
$\rho^{T_A}$ obtained from the partial transposition of $\rho$ in
Fock space, then it can be shown that $|\tilde \phi \rangle =
\sum_{nm}c_{nm} \sum_k \langle n |\tilde k \rangle | \tilde k
\rangle \sum_j \langle \tilde j|m \rangle | \tilde j\rangle$ is the
corresponding eigenvector of $\rho^{T_A}$ obtained from the partial
transposition of $\rho$ in the new bases (labeled by tilde), with
the same eigenvalue. Therefore $| \phi \rangle$ and $|\tilde \phi
\rangle$ are related by a direct bases transformation if $\langle n
|\tilde k \rangle = \langle \tilde k |n \rangle$, i.e., when
$\langle n |\tilde k \rangle$ is real.
\end{references}
\end{document}